\documentstyle[12pt,a4,openbib]{article}
\begin{document}

\newtheorem{theorem}{Theorem}
\newtheorem{lemma}{Lemma}
\newtheorem{propn}{Proposition}
\newtheorem{defn}{Definition}

\newcommand{\be}{\begin{eqnarray}}
\newcommand{\ee}{\end{eqnarray}}
\newcommand{\bes}{\begin{eqnarray*}}
\newcommand{\ees}{\end{eqnarray*}}
\newcommand{\beqn}{\begin{equation}}
\newcommand{\eeqn}{\end{equation}}

\newcommand\Pa{Painlev\'e}
\newcommand\tilu{\tilde u}
\newcommand\tils{\tilde s}
\newcommand\tilt{\tilde t}
\newcommand\tilx{\tilde x}
\newcommand\tilf{\tilde f}
\newcommand\mys{{\cal S}}
\newcommand\calF{{\cal F}}
\newcommand\calFp{{\cal F}\,'}
\newcommand\calL{{\cal L}}
\newcommand\calO{{\cal O}}
\newcommand\calU{{\cal U}}
\newcommand\Complex{{\rm\kern.24em \vrule
width.02em height1.4ex
     depth-.05ex \kern-.26em C}
}

\title{
Complex Blow-Up in Burgers' Equation: an Iterative
Approach}
\author{Nalini Joshi
         and
         Johannes A. Petersen\\
         {\it  School of Mathematics}\\
         {\it University of New South Wales}\\
         {\it  Sydney NSW 2052, Australia}
}
\maketitle
\begin{center}
{\bf Abstract}
\end{center}
We show that for a given holomorphic noncharacteristic surface
$\mys\in\Complex^2$, and a given holomorphic function on $\mys$,
there exists a unique meromorphic solution of
Burgers' equation which blows up on $\mys$.
This proves the convergence of the formal Laurent series
expansion found by the Painlev\'e test.
The method used is an adaptation of Nirenberg's iterative
proof of the abstract Cauchy-Kowalevski theorem.

\noindent {AMS Numbers: 35Q53, 35R20, 47H10\hfil}
\section{Introduction}
A partial differential equation (PDE) is said to have the \Pa\ property if
all  solutions are single-valued around all noncharacteristic holomorphic
movable singularity manifolds, where movable means that the manifold's
location depends on initial conditions. In
practice, a necessary condition of the property is usually checked through
formal power series expansion (see \cite{wtc:jmp}). Here we
show, through an iterative method in $\Complex^2$, that such series
converge for Burgers' equation
\begin{equation}
u_t+uu_x=u_{xx}.
\label{burgers}
\end{equation}

The \Pa\ property has become a widely used
indicator for integrability (see
\cite{ars:cim,ars:jmp}),
meaning exact solvability through the
inverse scattering method \cite{as:siam,ac:cup} or
linearizability through a transformation of variables.
Burgers' equation
is regarded as integrable
because it can be linearized (to the Heat equation)
by the Cole-Hopf transformation
\cite{hopf:cmap,cole:qam}. Hence, according to Ablowitz {\it et al}
\cite{ars:cim,ars:jmp} it should possess the Painlev\'e
property. To check that it does, Weiss {\it et al}
\cite{wtc:jmp} proposed that one should formally expand all solutions
around an arbitrary noncharacteristic singularity manifold given
by $\Phi(x,t)=0$ in a power series with a leading term
\begin{equation}
u=\sum_{n=0}^{\infty} u_n(x,t)\Phi^{n+\alpha},
\label{wtc}
\end{equation}
where $\alpha$ is to be found.

The expansion may be simplified by using the noncharacteristic nature of
the singularity manifold $\Phi=0$ which implies $\Phi_x\not=0$. By the
implicit function theorem (rescaling $\Phi$ if necessary) we have
\beqn
\Phi = x-\xi(t),
\eeqn
near $\Phi=0$, where $\xi(t)$ is an arbitrary function of $t$.
Replacing
$x$ by $\xi(t)+\Phi$ throughout the series (\ref{wtc})
we get a series in
powers of $x-\xi(t)$ with coefficients $u_n$ that
are functions of $t$ alone. Formal
expansion then shows that
$\alpha=-1$ and that the coefficient $u_2(t)$ is arbitrary. Hence the series
Eqn(\ref{wtc}) formally represents a meromorphic general solution
described by two arbitrary functions of one variable, namely $\xi(t)$ and
$u_2(t)$, near the singularity manifold.

Although widely used, there are two obvious deficiencies in this
procedure. First, convergence is ignored. Second, the procedure yields
only necessary consequences of the \Pa\ property and
makes no statement about whether these are
sufficient.

In this paper, we overcome the first deficiency. Our
aim is to develop a method that will generalize to all integrable
PDEs. Here, we present a method
that does generalize. An announcement of its
generalization to the Korteweg-deVries equation was made in
\cite{njjap:nonlin}.
Although Burgers' equation may be solved through the Heat equation, we present
the details of our method for Burgers' equation here because of its
value as a more transparent nonlinear example than the Korteweg-deVries
equation.

The method we use is a generalization of one
given for the \Pa\ equations
(six classical  nonlinear second-order
ODEs) by Joshi and Kruskal \cite{njmdk:direct}.
They showed that each \Pa\ equation could
be recast as an integral equation suitable for iteration near movable
singularities. Furthermore, the iteration of this equation has a fixed point
which gives a meromorphic solution in a neighbourhood of each movable
singularity.

In Section 2, we recast Burgers' equation as an integral equation that is
suitable for  iteration near a movable singularity to prove the following
theorem.
\begin{theorem}
Let $\mys$ be a holomorphic surface in $\Complex^2$ given
by $\{ t=\xi (x)\}$. Then locally there exists a solution of Burgers'
equation
\beqn
u_x+uu_t=u_{tt},
\label{uburgers}
\eeqn
which has the form
\beqn
u(t,x)=-\,{2\over t-\xi(x)}+h(t,x),
\eeqn
near $\mys$ where $h(t,x)$ is holomorphic. Moreover,
\beqn
\lim_{t\to\xi(x)}\{u_t(t,x)-{1\over 2}{\bigl(u(t,x)-\xi'(x)\bigr)}^2\}
\eeqn
is a holomorphic function of $x$, which can be given arbitrarily
in advance.
\end{theorem}
\noindent Note that
in keeping with the PDE literature, we have taken Burgers'
equation to be given by Eqn(\ref{uburgers}). That is,
the roles of $t$ and $x$ have
been interchanged. Also, note that throughout the paper, $(t,x)$ refers
to a point in
$\Complex^2$.

Our proof was influenced by the iteration proof of the abstract
Cauchy-Kowalevski theorem  given by Nirenberg \cite{nir:ack}.
After the completion of our work, we learnt of a different approach
developed by Kichenassamy and Littman \cite{kl:cpde1} for nonlinear
Klein-Gordon equations.
\section{Proof of the Theorem}
In this section, we convert Burgers'
equation to an integral equation suitable
for iteration near $\mys$, and prove the theorem stated above.

Let $f(x)$ be any analytic function. We begin by fixing our notation. Assume
(without loss of  generality) that the origin lies on \mys . Let $D$ be an
open neighbourhood  of the origin in $\hbox{\Complex}^2$ where
$$
\tilf(t,x):= t\xi''(x)+f(x)
$$
is holomorphic. We can straighten the surface $\mys$ locally into
the $x$-plane $\{t=0\}$ by using a biholomorphism $
(t,x)\mapsto (t-\xi(x), x) =: (\tilt, \tilx)$, $u(x, t)\mapsto\tilu(\tilt,
\tilx)$. Notice that this changes Burgers' equation into
\beqn
\tilu_{\tilt\tilt} = \bigl(\tilde u -\xi'(x)\bigr)\tilde u_{\tilt}
 + \tilde u_{\tilx}.
\label{nburgers}
\eeqn
It is sufficient to find a solution $\tilu$
having the form
$$
\tilu =-\,{2\over\tilt}+\tilde h,
$$
where $\tilde h$ is holomorphic such that
$$
\lim_{\tilt\to 0}\{\tilu_{\tilt}-{1\over 2}{\bigl(\tilu -\xi'(x)\bigr)}^2\}
=f(x).
$$
In the following, we will assume that $\mys$
is already locally given by the
plane $\{t=0\}$. So Burgers' equation will be assumed to be
Eqn(\ref{nburgers}).

To obtain a suitable integral equation, integrate Eqn(\ref{nburgers})
as though
only the dominant terms
i.e. $\tilu_{\tilt\tilt}$, $\tilde u\tilde u_{\tilt}$, were present.
Then, dropping the tildes, we get
\beqn
u_t={1\over 2}\left(u-\xi'(x)\right)^2+\int_0^tdt\,u_x+f(x).
\label{integral1}
\eeqn
Change variables to the reciprocal
\beqn
U={1\over u-\xi'(x)}.
\label{U}
\eeqn
Then if $U$ does not vanish in some neighbourhood off the $x$-plane,
Eqn(\ref{integral1}) gives
\beqn
-U_t=FU
\label{Ut}
\eeqn
where
\beqn
FU(t):={1\over 2}+U(t)^2\left(\int_0^t\partial_x\bigl({1\over U(\tau)}
         \bigr)\,d\tau + \tilf(t,x)\right),
\label{F}
\eeqn
is well defined. Integrate Eqn(\ref{Ut}) once more to get
\beqn
U=\calF U,
\label{fixedpt}
\eeqn
where
\beqn
\calF := -\int_0^t\,FU(\tau )\, d\tau .
\label{calF}
\eeqn
Conversely, if we find a fixed point $U$ of the operator $\calF$
then the corresponding $\tilu :=1/U+\xi'(t)$ will solve
Eqn(\ref{nburgers}).

We will study the iteration of the
operator $\calF$ for functions $U$
of the form
$$
U=-{t\over 2} + O(|t|^2).
$$
Note that substitution of such a function into Eqn(\ref{fixedpt})
reproduces
a function of the same form.

Let $\calO_0$ be an open
neighbourhood of the origin in
$\Complex$ and $d>0$ be a real number. Then for $0< s\le 1$, define
$$
\calO_s := \{x|\hbox{ dist}\,(x, \calO_0)< sd\}.
$$
We assume $\calO_0$ and $d$ small enough that $D$ contains the disk
$\{0\}\times\calO_1$.
Define for any number $a>0$
\beqn
D_a:=\{(t,x)\in\Complex^2\,\bigm|\, \exists\ 0\le s\le 1\hbox{ s.t. }
|t|< a(1-s) \hbox{ and }x\in\calO_s\},
\eeqn
and assume $a$ small enough that
$D_a$ is a subset of $D$.
For any real number $K$ and integer $n$, let
\bes
O^n_K(D_a):=\{U:D_a&\rightarrow&\Complex\,|\, U \hbox{ is holomorphic and }\\
               & &\forall (t,x)\in D_a\,\ |U(t,x)|\le K|t|^n\}.
\ees
These spaces denote remainder terms in Taylor expansions. Their union will
be written as
$$
O^n(D_a) := \bigcup_K O^n_K(D_a).
$$
The function spaces in which we will work are given by
$$
B^K_a:=\{U:D_a\rightarrow\Complex\,\bigm|\, U \hbox{ is holomorphic and }
                U=-t/2 +O^2_K(D_a)\},
$$
equipped with the {\sl sup}-norm on $D_a$. Our aim is to find a number $a>0$
and
a holomorphic function $U\in B^1_a$ that solves the fixed point equation
Eqn(\ref{fixedpt}).
We accomplish this by showing that the sequence $\{U_n\}$ of
iterates defined recursively by
\[
U_0=-\,{t\over 2}, \quad U_{n+1}=\calF U_n,
\]
converges to the desired fixed point of $\calF$.

In general, our (Newton) iteration method consists of two stages,
one linear, and the other
nonlinear, where the linear part is given
by the iteration of the Fr\'echet derivative of $\calF$. However,
for Burgers' equation it is sufficient to take this derivative
to be zero. (This is not the case for
the Korteweg-deVries equation.)

Our proof relies on
the following lemmas.
Proofs of these are given in subsections at the end of this
section.

\begin{lemma}
Suppose $a$ and $K$ are given positive numbers
such that\hfill\break $a<\min\{1/6, 1/(6K)\}$. If $U\in B_a^K$ then there is a
holomorphic function $g:D_a\rightarrow\Complex$ such that
$$
U(t,x)\bigl(-\,{2\over t}+ g(t,x)\bigr)=1
$$
wherever $U\not=0$. Moreover, $|g|$ is bounded by $6K$.
\end{lemma}
\begin{lemma}
Let $n\ge 0$, $a>0$ and $K>0$ be given numbers,
$0<\epsilon<1$, and $0<a^*\le a(1-\epsilon)$. Assume that the
holomorphic function $g:D_a\rightarrow\Complex$ satisfies for all
$(t,x)\in D_a$
$$
|g(t,x)|\le\,K|t|^n.
$$
Then for all $(t,x)\in D_{a^*}$ we get
$$
\Bigl|\int_0^t\partial_xg(\tau,x)\,d\tau\Bigr|\le
                {a\over d}\ln\left({1\over\epsilon}\right)K|t|^n.
$$
\end{lemma}
\begin{lemma}
Let $n\ge 1$, and suppose $a$, $a^*$, $K$, $L$ are
given positive numbers which satisfy
\bes
K&\ge&\sup_D\{|\tilf|, |f|\},\quad a<\min\{{1/6}, {1/(6K)}, d\},\\
& &a^n<{1/(12L)},\quad a^*\le a(1-2^{-(n+2)}).
\ees
Assume that $U_1$, $U_2$ are elements of $B^K_a$ and their difference
satisfies
$$v:= U_2 - U_1\in O^{n+1}_L(D_a).$$
Then we have
$$
\calF U_2 -\calF U_1\in O^{n+2}_{10L}(D_{a^*}).
$$
\end{lemma}
The last lemma is the key to the proof of our theorem. We will
apply it to the sequence $\{U_n\}$ in the sense that if
the iterates $U_{n-1}$ and $U_n$ already agree up to order
$n+1$, then the next pair of iterates $U_n$ and $U_{n+1}$
will agree up to order $n+2$.

\noindent{\bf Proof of the Theorem:} Let
$$
K\ge\sup_D\{|\tilf (t,x)|, |f(t,x)|\}.
$$
Note that this implies
$$
\sup_D|t\xi''(x)|\le 2K.
$$
Now assume
$$
0< a_0<\min\left\{{1/11}, {1/(11K)}, d\right\}.
$$
(As always, $a_0$ is assumed to be sufficiently small such that
$D_{a_0}\subset D$.) Moreover, define a sequence $\{a_n\}$ recursively
by
$$
a_{n}:=a_{n-1}(1-2^{-(n+2)}).
$$
We start the iteration in $D_{a_0}$ with
$$
U_0:=-\,{t\over 2},\quad
v_0:=\calF U_0-U_0.
$$
Note that for all $(t,x)\in D_{a_0}$, $v_0$ is bounded by
\bes
|v_0|&=&\Bigl|\int_0^t\bigl(t\xi''(x)+f(x)\bigr){t^2\over 4}\Bigr|\\
     &\le&{5K|t|^3\over 24}\ \le{Ka_0|t|^2\over 4}.
\ees
Let $L=Ka_0/4\,(\le 1/44)$. We have $U_0\in B_{a_0}^0$, $v_0\in
O^2_L(D_{a_0})$.

Now for the inductive step, suppose we have
$$
U_{n-1}\in B^1_{a_{n-1}},\quad
v_{n-1}\in O^{n+1}_{L_{n-1}}(D_{a_{n-1}}),
$$
where $L_{n-1}=10^{n-1}L$. Define
\beqn
U_{n}:=U_{n-1}+v_{n-1} = U_0+\sum_{j=0}^{n-1} v_j.
\label{Un}
\eeqn
We now show that $U_{n}\in B^1_{a_{n}}$. First
note that
$D_{a_n}\subset D_{a_{j}}$, $j=0,\ldots , n-1$. The induction
hypothesis gives $v_j\in O^{j+2}_{L_j}(D_{a_n})$. That is, for all
$0\le j\le n-1$
\beqn
|v_j|\le\left({10\over 11}\right)^jL\,|t|^2.
\label{vj}
\eeqn
Hence we get
$$
|U_{n}-U_0|\le 11 L|t|^2,
$$
which implies that $U_{n}\in B^1_{a_{n}}$ because $11 L<1$.

Now we apply Lemma 3 (with $a$, $a^*$, $L$, $v$ replaced by
$a_{n-1}$, $a_n$, $L_{n-1}$, $v_{n-1}$ respectively) to get an
estimate on $v_n$. Note that the hypothesis
\[a_{n-1}^n\le {1\over 12 L_{n-1}}\]
follows from $a_{n-1}\le a_0\le\min\bigl(1/(11), 1/(11K)\bigr)$ and
the definition of $L$.
Hence we get
\[v_n\in O^{n+2}_{L_n}(D_{a_n}).\]

The sequence $\{U_n\}$, $n=0, 1, 2, \ldots$, produced by the iteration
process above, is contained in $B^1_a$ where
\beqn
a:=\lim_{n\to\infty}a_n = a_0\prod_{n=0}^\infty\Bigl(1-2^{-(n+2)}\Bigr) > 0.
\eeqn
Consider now the limit of the sequence $\{U_n\}$. From Equations
(\ref{Un})(at ${n+1}$) and (\ref{vj}), we have
\beqn
U_{n+1}-U_n=v_n\in O^2_{L(10/11)^n}(D_a).
\eeqn
Hence it follows that $\{U_n\}$ is a convergent sequence and that the limit
$$
U:=\lim_{n\to\infty}U_n
$$
lies in $B^1_a$. Writing $||\ ||$ for the {\sl sup}-norm on $D_a$, we get for
any positive integer $n$
\be
||\calF U-U||&=&||\calF U - \calF U_n + U_{n+1}-U||\\
             &\le&||\calF U-\calF U_n||+||U_{n+1}-U||.
\ee
So by continuity of $\calF$ we can conclude that a fixed point
for $\calF$
$$
\calF U=U
$$
exists in $B^1_a$.\hfill
\subsection{Proof of Lemma 1}
\noindent{\bf Proof:}
The number $a$, in the definition of $D_a$, was assumed to be sufficiently
small
that $U$ does not vanish off the $x$-plane. Let $(t,x)\in D_a$, with
$t\not=0$. Then we can define $g$ to be
$$
g(t,x):= {1\over U(t,x)} + {2\over t}.
$$
The bounds on $a$, $K$, and $|t|$ give $|U(t,x)|>|t|/3$. Therefore,
$$\left|{1\over tU(t,x)}\right|< {3\over |t|^2}$$
and so
$$|g(t,x)|=\left|{t+2U(t,x)\over tU(t,x)}\right|\le 6K.$$
Since $g$ is then bounded and holomorphic off the $x$-plane, by
Kistler's theorem (see Osgood [11]), it can be extended to all of
$D_a$.

\subsection{Proof of Lemma 2}
\noindent{\bf Proof:} Let $(t,x)\in D_a^*$, and $|\tau|<|t|$. We put
\be
s'&:=&1-{|t|\over a^*},\\
s(\tau)&=&1-{|\tau|\over a}.
\ee
Note that $s(\tau)>s'$ and $D_{a^*}\subset D_{a}$ by the assumed
properties of $a^*$, $\tau$ and $t$. Similarly, $x\in\calO_{s'}$,
$\{\tau\}\times
O_{s(\tau)}\subset D_a$. So we can apply the Cauchy estimate
$$
||\partial_xg(\tau ,.)||_{s'}
           \le{||g(\tau ,.)||_{s(\tau)}\over d(s(\tau)-s')}
$$
together with the hypothesis on $g$ to get
$$
\left|\int_0^t\partial_xg(\tau ,x)\,d\tau\right|
           \le K\left|\int_0^t{|\tau|^n\,d\tau\over d(s(\tau)-s')}\right|,
$$
because $x\in O_{s'}$. Using the substitution $\tau =rt$ we get
$$
{1\over s(\tau)-s'}\le {a^*\over |t|\bigl(1-(1-\epsilon)r\bigr)}
$$
and so
\be
\Bigl|\int_0^t\partial_xg(\tau ,x)\,d\tau\Bigr|
           &\le& {a^*\over d}K|t|^n\int_0^1{r^n\,dr\over
                     \bigl(1-(1-\epsilon)r\bigr)}\\
           &\le& {a^*\over d}K|t|^n{\ln(1/\epsilon)\over 1-\epsilon}.
\ee

\subsection{Proof of Lemma 3}
\noindent{\bf Proof:} In the following, we will drop references to $D_a$,
i.e. $O^n_K(D_a)$ will be written as $O^n_K$. Also wherever convenient, we
will denote an element of $O_K^n$ by the set symbol $O_K^n$ itself. Since
$U_1\in B^K_a$ we have, for all $(t, x)\in D_a$,
$$
{|t|\over 2}-K|t|^2\le |U_1|\le {|t|\over
2}+K|t|^2.
$$
By the given hypotheses on $a$, $K$, and $t$, we then get
$$
{|t|\over 3}\le |U_1|\le {2|t|\over 3},
$$
which implies ${v/U_1}\in O^n_{3L}$. Now using $a^n\le 1/(12L)$, we have
$$
{v\over U_1}(1+3La^n+9L^2a^{2n}+27L^3a^{3n}+\ldots )\in O^n_{4L},
$$
implying that
$$
{1\over U_2}={1\over U_1} {1\over 1+v/U_1}={1\over U_1} (1+O^n_{4L}).
$$
By similar calculations, we get
$$v^2\in O^{n+2}_{L},\quad U_2^2=U^2_1+O^{n+2}_{3L}.$$
So by using Lemma 1, we get
$$
FU_2-FU_1=O^2_{4/9}\int^t_0\partial_xO^{n-1}_{12L}+O^{n+2}_{3L}
         \left(\int^t_0\partial_xO^0_{6K}+\tilf\right).
$$
So far all estimates have been obtained in $D_a$. Now we apply Lemma 2, and
thereby restrict our domain to $D_{a^*}$, to estimate the terms
differentiated with respect to $x$ in the above equations. To apply Lemma 2,
note that $\epsilon = 2^{-(n+2)}$ and that $a<d$. Then for any given integer
$N\ge 0$ and given $k>0$, we get
$$
\int^t_0\partial_xO^{N}_{k}(D_a)\le (n+2)O^{N}_k(D_{a^*}).
$$
Recalling that $|\tilf|<K$ and using the definition of $\calF U$, we get
\be
\calF U_2-\calF U_1&=&\int^0_t\Bigl\{FU_2(\tau)-FU_1(\tau)\Bigr\}d\tau\\
&=&\int^0_t\Bigl\{(n+2)O^{n+1}_{16L/3}+(n+2)O^{n+2}_{18KL}
        +O^{n+2}_{3KL}
            \Bigr\}d\tau\\
 &=&\int^0_t\Bigl\{(n+2)O^{n+1}_{25L/3}+O^{n+1}_L\Bigr\}d\tau
\ee
where the last line is obtained by using $aK<1/6$ and the integrands after the
first line are evaluated on $D_{a^*}$. Integration gives the desired result
$$\calF U_2-\calF U_1\in O^{n+2}_{10L}(D_{a^*}).$$

\section{Acknowledgements}
The research reported here was supported by the Australian Research Council.
NJ would also like to thank Martin D. Kruskal for early conversations on the
iterative proof of the Cauchy-Kowalevsky theorem and Fran\c cois Treves for
informing us about Ref. \cite{nir:ack}.

\bibliography{refs}
\bibliographystyle{plain}
\end{document}